\documentclass[10pt,conference]{IEEEtran}
\usepackage{amsmath,amssymb,amsfonts}
\usepackage[linesnumbered,ruled,vlined]{algorithm2e}
\usepackage{epsfig}
\usepackage{epstopdf}
\usepackage{multirow}
\usepackage{hhline}
\usepackage{verbatim}
\usepackage{algorithmicx}
\usepackage{lipsum}
\usepackage{algpseudocode}
\usepackage{graphicx,subfigure}
\usepackage{enumitem}
\usepackage{graphicx}
\usepackage{textcomp}
\usepackage{algcompatible}
\usepackage{relsize}
\usepackage{multicol}
\usepackage{multirow}
\usepackage{xcolor}
\usepackage{caption}
\usepackage{braket}
\usepackage[c3]{optidef}
\newtheorem{property}{Property}

\usepackage[ruled,vlined]{algorithm2e}

\newcommand{\Qc}{{\cal{Q}}}

\newcommand{\Kc}{{\cal{K}}}
\newcommand{\Ec}{{\cal{E}}}
\newcommand{\Gc}{{\cal{G}}}
\newcommand{\Hc}{{\cal{H}}}

\newcommand{\Nc}{{\cal{N}}}

\newcommand{\Vc}{{\cal{V}}}

\newcommand{\Gct}{\widetilde{\cal{G}}}

\title{Energy-Efficient Dynamic Training and Inference for GNN-Based Network Modeling\vspace*{-0.15in}}%
\makeatletter
\newcommand{\linebreakand}{%
  \end{@IEEEauthorhalign}
  \hfill\mbox{}\par
  \mbox{}\hfill\begin{@IEEEauthorhalign}
}
\makeatother

\author{
  \IEEEauthorblockN{Chetna Singhal}
  \IEEEauthorblockA{\textit{Inria, Univ Rennes, CNRS, IRISA} \\
  Rennes, France \\
    chetna.iitd@gmail.com}\vspace*{-1cm}
  \and
  \IEEEauthorblockN{Yassine Hadjadj-Aoul}
  \IEEEauthorblockA{\textit{Univ Rennes, Inria, CNRS, IRISA} \\
    Rennes, France \\
yassine.hadjadj-aoul@irisa.fr}
\vspace*{-1cm}} 

\begin{document}
\setlength{\columnsep}{0.248in}

\maketitle 
\begin{abstract}
Efficient network modeling is essential for resource optimization and network planning in next-generation large-scale complex networks. Traditional approaches, such as queuing theory-based modeling and packet-based simulators, can be inefficient due to the assumption made and the computational expense, respectively. To address these challenges, we propose an innovative energy-efficient dynamic orchestration of Graph Neural Networks (GNN) based model training and inference framework for context-aware network modeling and predictions. We have developed a low-complexity solution framework, QAG, that is a Quantum approximation optimization (QAO) algorithm for Adaptive orchestration of GNN-based network modeling. We leverage the tripartite graph model to represent a multi-application system with many compute nodes. Thereafter, we apply the constrained graph-cutting using QAO to find the feasible energy-efficient configurations of the GNN-based model and deploying them on the available compute nodes to meet the network modeling application requirements. The proposed QAG scheme closely matches the optimum and offers atleast a 50\% energy saving while meeting the application requirements with 60\% lower churn-rate.
\end{abstract}

\begin{IEEEkeywords}
Network modeling, Adaptive GNN Training and Inference, Machine learning orchestrator, Energy-efficiency, Mobile-edge-cloud continuum\vspace{-2mm}
\end{IEEEkeywords}

\captionsetup{font=footnotesize}
\section{Introduction}
\vspace{-0.15cm}
Accurate network modeling is essential for applications like network digital twins (NDT), resource optimization, and traffic flow control in complex next-generation networks. While packet-level simulators can provide high accuracy, they come with significant computational costs. Network simulators like ns3 and Omnet++ may require considerably long simulation times (several hours) depending on the network and the traffic flow characteristics~\cite{ns1,routenetFermi}. Furthermore, methods such as the Markov model, queuing theory, and network simulators have been employed for these tasks, but they introduce assumption-based limitations and increased computational costs.

Recently, machine learning has been applied for efficient data-driven network modeling. Graph Neural Networks (GNN) can learn complex non-linear behavior in networks and predict QoS parameters with a level of accuracy similar to computationally expensive packet-level simulators~\cite{routenetFermi}. GNNs have the potential for accurate network delay and jitter prediction in future networks, even with complex topologies~\cite{Transnet,xnet}. However, the GNN model training can be computationally expensive and time-consuming~\cite{xnet}. The machine learning inference performance depends highly on the model architecture and computation platform~\cite{fin_infocom2024,secon2024}. The existing works do not deal with application-awareness or dynamic network modeling using GNNs, which are essential for energy-efficient orchestration in the mobile-edge-cloud continuum network. 

\begin{figure}[!htb]
    \centering
        \vspace{-0.05in}
    \includegraphics[width=2.6in]{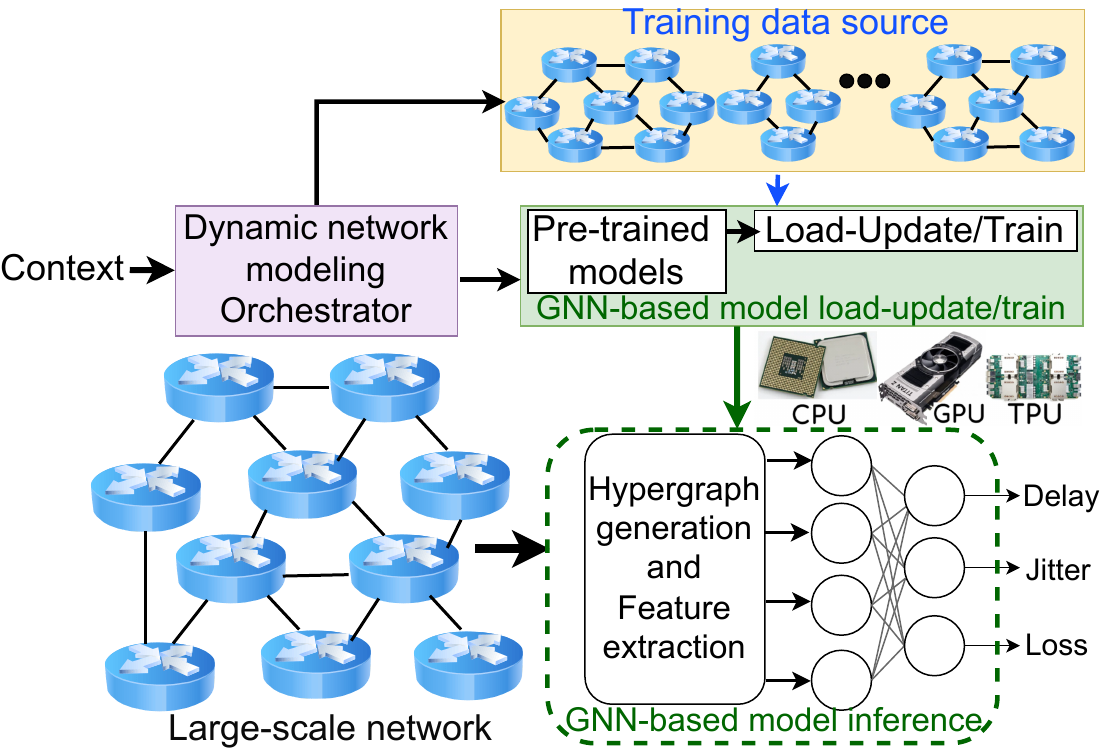}
    \vspace{-0.05in}
    \caption{Dynamic orchestration of GNN-based model training and inference for network modeling and prediction.}
    \vspace{-0.1in}
    \label{fig:intro}
    \vspace{-2mm}
\end{figure}

Our work lies at the intersection of two major fields, namely, machine learning for network modeling and dynamic machine learning. In this paper, we have developed an energy-efficient dynamic GNN orchestration framework for efficient network modeling, as illustrated in Fig.~\ref{fig:intro}. Network modeling applications require delay, jitter, and packet-loss predictions for large-scale networks with a required level of inference latency and quality (loss). The {\em{orchestrator}} uses such context information to dynamically model large-scale networks. The mobile-edge-cloud continuum system consists of compute nodes, such as, Central Processing Unit (CPU), Graphical Processing Unit (GPU), and Tensor Processing Unit (TPU). These nodes handle the GNN model training/update and inference for heterogeneous applications. We have developed a dynamic network modeling orchestrator that decides whether to deploy a pre-trained model or update/train it using a smaller-network data source such that the application inference requirements are met. In doing so, it also determines the compute node over which the GNN based model executes.
  \vspace{-1mm}

The main contributions of this work are given as follows:
\begin{enumerate}[itemsep=0mm,leftmargin=5mm]
\item We model the energy consumption and latency performance of the training-update-inference task of the GNN-based network modeling and optimization applications.
\item We minimize the energy consumption at the system level to deploy network modeling applications, while meeting their quality (loss) and latency requirements.
\item We design an energy-efficient orchestrator that dynamically selects the extent of train-update-infer function to meet the application requirements and system constraints.
\item We model the system using a tripartite graph and develop a low-complexity solution, QAG, using the Quantum approximation optimization to solve the above problem.
\end{enumerate}

 
\section{System Model}
\vspace{-2mm}
We consider a computing framework composed of CPUs, GPUs, and TPUs that dynamically trains GNN models to predict QoS parameters. The objective of the overall system is to support a set of such network modeling applications whose central component is a set of GNN models that are trained (with updates) using the small-network information and thereafter perform predictions using the large-scale network information to meet the required application requirements. 

The edge of the network infrastructure hosts a network modeling orchestrator that possesses knowledge of the applications and capabilities of the computing nodes' resources, network datasets, and configurations. 
Formally, the main elements defining the system are:\\
$\bullet$ A set $\boldsymbol{\sigma}$ of GNN model configurations indexed with $\sigma{\in}\{1,\ldots,S\}$, composed of a predefined set of stages in the message-passing architecture and a data source. The data sources contain small-scale network information that serves as input to the training-update module and facilitates the GNN model for network predictions on the large-scale network for applications. Each configuration is defined as a tuple $\sigma=\{\texttt{data\_source},\texttt{epochs},\texttt{steps\_per\_epoch},\psi\}$,
\begin{equation}
\label{eq:mode}
\text{\!\!where, }\psi\!=\!\begin{cases}
			0, & \!\!\text{\!\!load pre-trained model \& infer}\\
            1, & \!\!\!\text{\!\!load pre-trained model, update, \& infer}\\
		 \end{cases}.
\end{equation}
$\bullet$ A set $\Hc$ of applications $h{\in}\{1,\ldots,H\}$. Each application has specific requirements defined as the target inference quality (e.g., target loss) $\ell^h_{max}$, and the maximum inference latency $\tau^h_{max}$. In the following, we refer to the applications and the GNN representing their essential components interchangeably.\\
$\bullet$ A set $\Nc$ of computationally-capable nodes, including (i) GPUs, (ii) CPUs, and (iii) TPUs. 
    
Given an application (context) $h$, the orchestrator determines which configuration of the GNN model should be used, and to what extent should it train-update-infer. It assigns these models to the computation nodes (CPUs, GPUs, and TPUs) in such a way that the application inference requirements are fulfilled. Note that a computing node may be allocated zero, one, or multiple GNN models, which are all executed.

We represent the overall system by means of an \emph{application-load-resource weighted complete tripartite graph} model~\cite{bipartite1} with three sets of vertices, i.e., \emph{application}, \emph{GNN model configurations (data source and architecture)}, and \emph{computing nodes}. We establish a relationship between the applications' requirements (quality and latency) in $\Hc$, the compute load of configurations in $\boldsymbol{\sigma}$, and the resources made available by the heterogeneous computing nodes in $\Nc$. It matches resources handled by the system, i.e., offered by the computing nodes and the computing load required by the applications.

\begin{figure}
  \includegraphics[width=3.5in]{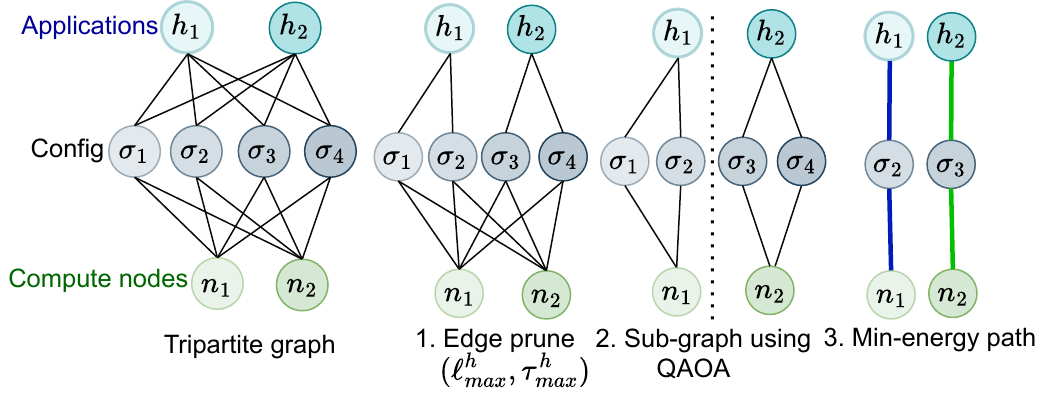}
  \vspace{-7mm}
    \caption{Application-load-resource tripartite graph model. This small-scale example shows orchestration steps of two applications (blue and green edges) in a system with four configs. (combination of two GNN architectures and two data sources), and two computing nodes (CPU and GPU).  \vspace{-3mm}}
    \label{f:example}
\end{figure}
\begin{figure}
\centering
\scriptsize
\parbox{.35\linewidth}{
\begin{tabular}{ |c|c|c| } 
\hline
[s,t]& $c^h$ & $\ell^h$  \\
\hline
$h_1,\sigma_1$\!\!& 50 & 15 \\ \hline
$h_1,\sigma_2$\!\!& 30 & 25 \\ \hline
$h_1,\sigma_3$\!\!& 20 & 40 \\ \hline
$h_1,\sigma_4$\!\!& 10 & 50 \\  \hline
$h_2,\sigma_1$\!\!& 100 & 65 \\ \hline
$h_2,\sigma_2$\!\!& 80 & 75 \\ \hline
$h_2,\sigma_3$\!\!& 70 & 20 \\ \hline
$h_2,\sigma_4$\!\!& 60 & 30 \\  \hline
\end{tabular}
\vspace{0.2in}
}
\parbox{.35\linewidth}{
\centering
\scriptsize
\begin{tabular}{ |c|c|c| } 
\hline
[s,t]& $c^1,c^2$ & $\ell^1,\ell^2$  \\
\hline
$\sigma_1,n_1$\!\!& 50,100 & 15,65 \\ \hline
$\sigma_2,n_1$\!\!& 30,80 & 25,75 \\ \hline
$\sigma_3,n_1$\!\!& 20,70 & 40,20 \\ \hline
$\sigma_4,n_1$\!\!& 10,60 & 50,30 \\  \hline
$\sigma_1,n_2$\!\!& 50,100 & 15,65 \\ \hline
$\sigma_2,n_2$\!\!& 30,80 & 25,75 \\ \hline
$\sigma_3,n_2$\!\!& 20,70 & 40,20 \\ \hline
$\sigma_4,n_2$\!\!& 10,60 & 50,30 \\  \hline
\end{tabular}
\vspace{0.2in}
}
 \vspace{-5mm}
    \caption{Edge weight for application ($h\in\Hc$, $v_i\in\Vc_1$) to configuration ($\sigma\in\boldsymbol{\sigma}$, $v_j\in\Vc_2$) (left) and configuration  to compute node ($n\in\Nc$, $v_k\in\Vc_3$) vertices (right) in the tripartite graph.\label{f:weights}  \vspace*{-6mm}}
\end{figure}

We show the tripartite-graph representation for a small-scale scenario in Fig.\,\ref{f:example}. It consists of two applications, four possible GNN-based model configurations, and two compute nodes. 
More formally, we denote the graph, illustrated in Fig.\,\ref{f:example}, with $\Gc=\{\Vc,\Ec\}$ where $\Vc$ are the vertices and $\Ec$ are the edges. The vertices $V_1{\subset} \Vc$ correspond to the heterogeneous applications $h_i\in\Hc$. The vertices $V_2{\subset} \Vc$ correspond to the GNN model configurations $\sigma_j\in\boldsymbol{\sigma}$. The vertices $V_3{\subset}\Vc$ correspond to the system computing nodes $n_k\in\Nc$. A resource sharing setting is in place, where applications and corresponding selected configurations are assigned separate portions of computing resources on selected computing nodes.

The computational resource requirement (in operations) and inference quality (loss) associated with the edge connecting vertices $v_i$ and $v_j$  for application $h$ are indicated by $c^h(v_i,v_j)$ and $\ell^h(v_i,v_j)$, resp. We associate each edge in $\Ec$ with a multidimensional weight $[c^h(v_i,v_j), \ell^h(v_i,v_j)]$. The edge weights for the small-scale example (in Fig.~\ref{f:example}) are given in Fig.~\ref{f:weights}. 

We use the following indicator variable that denotes whether the orchestrator selects the corresponding vertices of the edge to be active in the system. 
\begin{equation}
{
\small
\chi^h(v_i,v_j)=\begin{cases}
			1, & \text{if $v_i$ and $v_j$ are active for application $h$}\\
            0, & \text{otherwise.}
		 \end{cases}
   }
\end{equation}
The set containing the indicator variable values for the tripartite graph is denoted as $\boldsymbol{\chi}$. 
The latency of the train-update-inference task for application $h\in\Hc$ is defined as:

{\small{\vspace{-3mm}
\begin{equation}
    \tau^h=\text{\hspace{-4mm}}\sum\limits_{\forall v_i,v_j\in\Vc, v_i\neq v_j}\text{\hspace{-4mm}}t^h(v_i,v_j)\cdot \chi^h(v_i,v_j),\, 
    t^h(v_i,v_j)=\frac{c^h(v_i,v_j)}{x^h_c(v_j)}
\end{equation}}
}
where, $x^h_c(v_j)$ is the allocated compute resources [in operations per second] to application $h$ at compute node $v_j\in V_3$. 
We denote the required resource by a configuration $v_i\in V_2$ for application $h$ deployed on compute node $v_j\in V_3$ as $c^h(v_i,v_j)$. The value of $x^h_c(v_j)=\infty,\,\forall v_j\in V_1\cup V_2$, i.e. there is no compute resource limit on the configuration vertices. 
Energy consumption\,[in J]\,for application $h$ is defined as: 

\begin{equation}
    E^h=\sum\limits_{\forall v_i,v_j\in\Vc, v_i\neq v_j}t^h(v_i,v_j)\cdot \chi^h(v_i,v_j)\cdot \xi(v_j)
\end{equation}
 where,  $\xi(v_j)$ [in W] is the  power consumption of the compute node $v_j\in V_3$ and $\xi(v_j)=0,\,\forall v_j\in V_1\cup V_2$.

\vspace{-1mm}
\section{Context-aware GNN-based Network Modeling\label{sec:system}}
\vspace{-2mm}
\subsection{Problem formulation}
\vspace{-1mm}
The optimization problem for the orchestrator is: 

\vspace{-2mm}
{\begingroup
\allowdisplaybreaks
 \begin{subequations}
 {
 \label{e:optimization}
 \smaller
 \vspace{-1mm}
\begin{align}
& \min_{{\boldsymbol{\chi}}}  \,\,\,  \sum_{h {\in} \Hc} E^h\label{P2}\\
 &\textrm{s.t.\,\,\,} \tau^h \leq \tau^h_{max} \label{C1_P2}\\
 \sum\limits_{\forall v_i\in V_1, \forall v_j\in V_2}\ell^h(v_i,&v_j)\cdot\chi^h(v_i,v_j)\leq \ell^{h}_{max}, \,~\forall h\in\Hc\label{C2_P2}\\
  \sum\limits_{\forall v_i\in V_1, \forall v_j\in V_2}\hspace{-7mm}\chi^h(v_i,v_j)\!\cdot c^h(v_i&,v_j)\leq \hspace{-6mm}\sum\limits_{\forall v_i\in V_2, \forall v_j\in\Vc_3}\hspace{-7mm}\chi^h(v_i,v_j)\!\cdot\!x_c^h(v_j)\,, \forall h\in\Hc\label{C6_P2}\\
 \sum\limits_{\forall v_i\in V_1, \forall v_j\in V_2}\!\!\!\!\!\!\!\chi^h(v_i,v_j)\!=&\!1, \sum\limits_{\forall v_i\in V_2, \forall v_j\in V_3}\!\!\!\!\!\!\chi^h(v_i,v_j)\!=\!1\label{C3_P2}, \,\,\forall h\in\Hc
\end{align}
}
\end{subequations}
\endgroup}
The objective \eqref{P2} of the orchestrator is to minimize the overall energy consumption in the system to deploy GNN  models, and facilitate the application set $\Hc$, subject to latency \eqref{C1_P2}, quality-loss \eqref{C2_P2}, and compute-resource constraints \eqref{C6_P2}. This is achieved by optimizing the values of the indicator variables in the set, $\boldsymbol{\chi}$, for the tripartite graph representation of the system. 
Specifically the constraint \eqref{C6_P2} ensures that the compute requirements of the configuration that is selected for an application $h\in\Hc$ is met by the resulting resource allocation at compute nodes. The constraint \eqref{C3_P2} ensures that only one configuration is selected for each application and only one compute node completely deploys the GNN based model for that application, i.e. there is no split in the train-update-inference of the model.

\noindent
\begin{property}
The optimization problem \eqref{P2} subject to constraints \eqref{C1_P2}--\eqref{C3_P2} is NP-hard.
\end{property}

\begin{IEEEproof}
We reduce a known NP-hard problem, the Steiner tree problem (STP)~\cite{hwang1992steiner}, to a {\em simplified} instance of our problem, defined in \eqref{P2}--\eqref{C3_P2}. 
Given an instance of the STP, we construct a corresponding instance of the problem by creating: a selected configuration for all the application vertices to connect except one; one GNN configuration, corresponding to the remaining vertex to connect; compute nodes for all configuration vertices in the STP instance; and only one of the compute nodes has enough capabilities to run the GNN configuration. Also, the deployment of the configurations on the compute nodes reproduces that of the STP instance, and one component of the weights in our problem instance is set to match the weights in the STP instance while all others are set to zero. Solving our problem to optimality thus yields an optimal solution to the STP instance. Hence, the two problems are equivalent. Since the reduction takes polynomial (linear) time (each edge and vertex of the STP instance is processed once) and the STP problem is NP-hard~\cite{hwang1992steiner}, the NP-hardness of our problem is proved.
\end{IEEEproof}

It is also worth noting that the instance of our problem created in the proof above is very simple, and that on top of being NP-hard, our problem is significantly {\em more} complex than STP. Hence, we propose below an algorithmic solution, leveraging a quantum approximate optimization approach applied to the graph representation that, efficiently and very conveniently, finds the feasible decisions that meet the application requirements and system constraints.


\vspace{-2mm}
\subsection{Solution framework: QAG}
We propose the solution framework QAG, Algorithm~\ref{a:qag}, that performs Quantum Approximate Optimization (QAO) for adaptive orchestration of GNN-based network modeling in the mobile-edge continuum systems. It consists of three functions, namely, \texttt{Edge\_prune}, \texttt{Sub\_graph\_using\_QAOA}, and \texttt{Min\_energy\_path}, performing the steps illustrated in the example that is shown in Fig.\,\ref{f:example}. The \texttt{Edge\_prune} function removes the infeasible edges (that violate the loss and latency requirement) from $\Gc$.  
\begin{algorithm}[htb]
\footnotesize
\caption{QAG: QAO for Adaptive orchestration of GNN-based network modeling}
\SetAlgoLined
\DontPrintSemicolon
\label{a:qag}
 \textbf{Input:} $\cal{G}$, $\tau^h_{max}$, $\ell^h_{max}$;
 \SetKwFunction{FMain}{Main}
  \SetKwFunction{FSum}{\!\!}
  \SetKwFunction{FSub}{\!\!}
  
\noindent\SetKwProg{Fn}{Function I: Edge\_prune}{:}{}
  \Fn{\FSum{$\Gc$,$\ell^h_{max}$,$\tau^h_{max}$}}{
  {\bf Remove infeasible edges}\;
      ${}$\hspace{0.4em}\For{\text{each vertex} $v\in \Vc$ and \text{each edge} $(v,v^\prime)\in\Ec$}{
      \If{$\ell^h(v,v^\prime)> \ell^h_{max}$ or $t^h(v,v^\prime)> \tau^h_{max}$}{
      Remove edge $(v,v^{\prime})$: $\Ec=\Ec\backslash(v,v^\prime)$\;
     }
     }
    \KwRet {\it{Updated graph: }$\Gc$}
  }

\noindent\SetKwProg{Fn}{Function II: Sub-graph\_using\_QAOA}{:}{}
  \Fn{\FSum{$\Gc$}}{
  {\bf 1. Create complement graph}\;
      ${}$\hspace{0.4em}\For{\text{each vertex} $v\in \Vc$}{
      Include $v$ in $\Vc^\prime$\;

      ${}$\hspace{0.0em}\For{\text{each vertex} $v^\prime\in\Vc$}{
      \If{$v\neq v^\prime$ and edge $(v,v^\prime)\notin \Ec$}{
      Include edge $(v,v^\prime)$ in $\Ec^\prime$\;
      
      }
      }
     }
       {\bf 2. QAOA($\Gc^\prime=\{\Vc^\prime,\Ec^\prime\}$)}\;
       Create QAOA circuit $\Qc$ using parameters in Table~\ref{t:sim}\;
       Simulate $\Qc$ using parameters in Table~\ref{t:sim}\;
       Optimize $\Qc$ to minimize objective function value\;
       {\bf 3. Create Sub-graphs}\;
       $\Vc_j=\{i|q_i=j-1, i\in[1,N]\}$, $j\in\{1,2\}$\;
       \eIf{$\Vc_j$ has a single application vertex}{Include $\Gc_j$ in $\Gct$}
       {Repeat lines 13-30 on $\Gc_j$}
    \KwRet {\it{Graph subsets: }}$\Gct=\{\Gc_1,\Gc_2,\ldots,\Gc_H\}$
  } 
\noindent\SetKwProg{Fn}{Function III: Min\_energy\_path}{:}{}
  \Fn{\FSum{$\Gct$}}{
    ${}$\hspace{0.4em}\For{\text{each application} $h\in \Hc$}{
      \!\!\!\!$\chi^h(v_i,v_j)\!=\!1, \text{ for all } (v_i,v_j) \text{ that } \min E^h, \forall v_i,v_j\in V_h$. 
     } 
    \KwRet $\boldsymbol{\chi}$ as QAG output
  } 
\end{algorithm}

In \texttt{Sub\_graph\_using\_QAOA} function we first define the complement of the tripartite graph $\Gc=\{\Vc,\Ec\}$ as $\Gc^\prime=\{\Vc^\prime,\Ec^\prime\}$, where $\Vc^\prime=\Vc$ and $\Ec^\prime=\Kc\backslash \Ec$, $\Kc$ consists of all two element subsets of $\Vc$, and $\Kc\backslash \Ec$ is the relative complement of $\Ec$ in $\Kc$.
Thereafter we apply the QAOA max-cut~\cite{qaoa_main,qaoa} on the complement graph $\Gc^\prime$ with the constraint that the subset graphs should have at least one vertex from each vertex set, $\Vc_1, \Vc_2$, and $\Vc_3$. 
The objective function to obtain best subsets using QAOA is defined as:
\begin{equation}
\min\limits_{q\in\{0,1\}^N}\sum\limits_{\substack{1\leq i \leq N, 1\leq j \leq N\\ i\neq j, (i,j)\in{\Ec^\prime}}}\text{\hspace{-0.2in}}-\big((q_i\cdot(1-q_j)+((1-q_i)\cdot q_j)\big)
\label{qaoa_eq}
\end{equation}
where, $N$ is the number of vertex in the graph. We optimize the $p$-layer QAOA circuit~\cite{qaoa_main} based on the above objective. QAOA circuit with the  Since we apply the QAOA max-cut on the complement graph, the above objective function minimization happens when the nodes connected by the edges in the complement graph are assigned to separate subsets. In terms of feasibility all the compute node vertices are equivalent due to pruning step and the resulting quantum states might reflect the subset graphs similarly. The highest probability quantum state $q=\ket{q_1q_2\ldots q_N},q_i\in\{0,1\},1\leq i \leq N$, gives the two subgraphs with vertices $V_j=\{i|(q_i=j-1)\},j\in\{1,2\}$. In case, the subgraph has more than one application vertex then further subgraphs are obtained from it using the above approach. Finally, the \texttt{Min\_energy\_graph} finds the minimum energy path for each application from its corresponding sub-graph and assigns the indicator variable ($\chi^h$) accordingly.

\begin{table}[!htb]
{\footnotesize
\centering
\caption{Parameter settings\vspace{-2mm}}
\label{t:sim}

    \begin{tabular}{|l|l||l|l|}
        \hline
       {\bf{\!\!Quantum parameter}}\!\!\!& {\bf{Value}} & \!\!\!{\bf{GNN-model parameter}}\!\!\!  & {\bf{Value}} \\    \hline 
{Quantum gate ($\gamma_0,\beta_0$) }& {(-1,-3)}  &
Learning rate & $10^{-3}$ \\ \hline
\!\!\!Number of layers $p$, shots\!\! & 2, 100 & Message-passing iter.& 8 
 \\ \hline
Number of iterations & 100 & Update rule & Adam   \\ \hline

\end{tabular}
}
\vspace{-5mm}
\end{table}
\begin{table}[!htb]
{\footnotesize
\centering
\caption{Application, scenario, predicted parameter, and data sources}
\vspace{-2mm}
\label{t:app}
\footnotesize
\begin{tabular}{|p{0.14in}|p{0.57in}|p{0.57in}|p{1.3in}|}
    \hline
   \!\!\!{\bf App.}&{\bf Scenario}&\!\!\!{\bf Predict\,par.}&{\bf Train/Test network data} \\ \hline
   $h_1$&\!\!Real-traffic &Delay &\!\!\!\!GERMANY50,\,NOBEL-GBN,\,GEANT,\,ABILENE\!\!\!\\ \hline
   $h_2$&\!\!\!Traffic\,models\!\!& Delay&GBN,\,GEANT,\, NSFNET\\ \hline
   $h_3$&\!\!\!Traffic\,models\!\!&Jitter&GBN, GEANT, NSFNET\\ \hline
   $h_4$&\!\!\!Traffic\,models\!\!& Packet-loss &GBN, GEANT, NSFNET\\ \hline
    $h_5$&\!\!\!Scheduling& Delay&GBN, GEANT, NSFNET\\\hline
    $h_6$&\!\!\!Scheduling& Jitter&GBN, GEANT, NSFNET\\\hline
    $h_7$&\!\!\!Scheduling& Packet-loss &GBN, GEANT, NSFNET\\\hline
    
\end{tabular}
\vspace{-5mm}
}
\end{table}
\begin{table}[!htb]
\centering
{\footnotesize
\caption{Compute nodes Power usage [W] and compute capacity [TOPS] \vspace{-2mm}}
\vspace{-5mm}
\label{t:compute}
\begin{center}
\begin{tabular}{|c|c|c|c|c|}
    \hline
   {\bf Compute}&{\bf Compute}&\multicolumn{2} {c|}{\bf Power\,[W]}&\!\!{\bf TOPS} \\ \cline{3-4}
    {\bf Node}&{\bf Type}& Idle & Max&   \\ \hline
    $n_1$&CPU  & 5&12 & 2 \\ \hline
    $n_2$&T4 GPU  & 36&70 & 80  \\ \hline
    $n_3$&TPU v2 &53&280 & 180 \\ \hline
\end{tabular}
\vspace{-2mm}
\end{center}
\vspace{-2mm}
}
\end{table}

The quantum approximation optimization algorithm offers a polynomial-level complexity~\cite{qaoa_wcnc}. The initialized QAO for max-cut with $p$ layers achieves a $O[poly(p)]$ complexity~\cite{qaoa_main}. For a $N$ node tripartite graph, the QAG algorithm with a two-layer architecture exhibits  $O[N+Hpoly(2)]$ complexity which is significantly lesser than the $O[N^N]$ complexity of the brute-force optimal exhaustive-search method.

\section{Performance Evaluation}
In this section, we describe the reference scenario and examine the impact of different GNN model configurations on latency, quality, and energy cost. Then, we introduce the benchmarks against which we compare QAG, and present the applications' performance in the reference system.
\subsection{Reference scenario}
\noindent\textbf{GNN-based model architecture. }
The network is defined by three main components: active flows in the network, queues at each output port of the network devices, and physical links. A 32 element hidden state vector encodes the initial features of these components by using a 2-layer fully-connected neural network with ReLU activation functions and 32 units each. This is followed by a three-stage message passing algorithm that combines and updates the state of the components over 8 iterations. Finally, the readout functions are implemented as 3-layer fully-connected neural networks with a ReLU activation function for the hidden layers and a linear one for the output layer. These are applied to the hidden states of the flow component across specific links to compute the delay, the jitter, and the packet loss. 

\noindent\textbf{Applications and model configurations. }
We consider the network modeling applications, listed in Table~\ref{t:app} that range across heterogeneous traffic and scheduling scenarios as well as prediction parameters such as delay, jitter, and packet-loss. The model configurations consist of the training/test data sources (\texttt{data\_source}) that are listed in Table~\ref{t:app}, a range of epochs \texttt{epochs}$\in[1,50]$, \texttt{steps\_per\_epoch}$\in [1,2000]$, and the model deployment mode $\psi\in\{0,1,2\}$ that is defined in~eqref{eq:mode}. Overall, this results in a large number of possible configurations ($>500k$) for each application.
\begin{figure}[!htb]
 \vspace{-3mm}
  \centering
\hspace{-0.1in}\subfigure[\vspace{-3mm}\label{f:loss_infer_app}]{
\includegraphics[width=0.4\columnwidth]{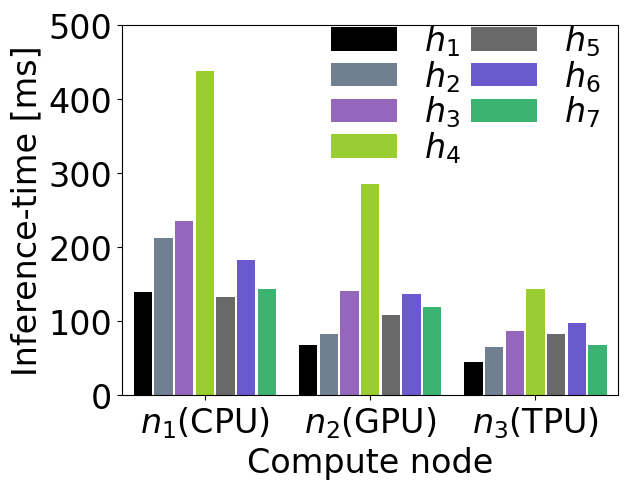}\vspace{-5mm}  
    }\hspace{1mm}  \subfigure[\vspace{-3mm} \label{f:infer_time_app}]{      \includegraphics[width=0.32\columnwidth]{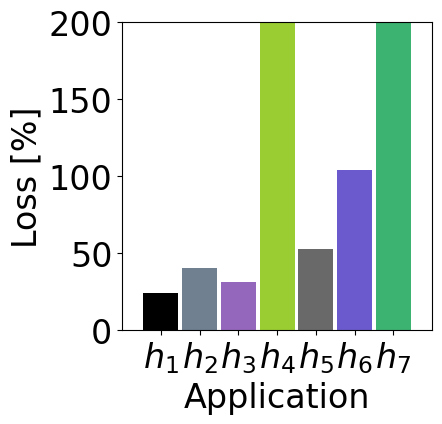}\vspace{-5mm}
         }
         \vspace{-3mm}
         \caption{(a) Inference-time/sample on compute nodes; and (b) Loss (MAPE) performance (config.: load-and-infer) for the applications listed in Table \ref{t:app}.\vspace{-5mm}}
 \label{f:res_infer}
 \end{figure}
 \vspace{-0.2cm}
\begin{figure}[!htb]
\vspace{-2mm}
  \centering
\hspace{-0.1in}\subfigure[\vspace{-3mm}\label{f:loss_app}]{
\includegraphics[width=0.49\columnwidth]{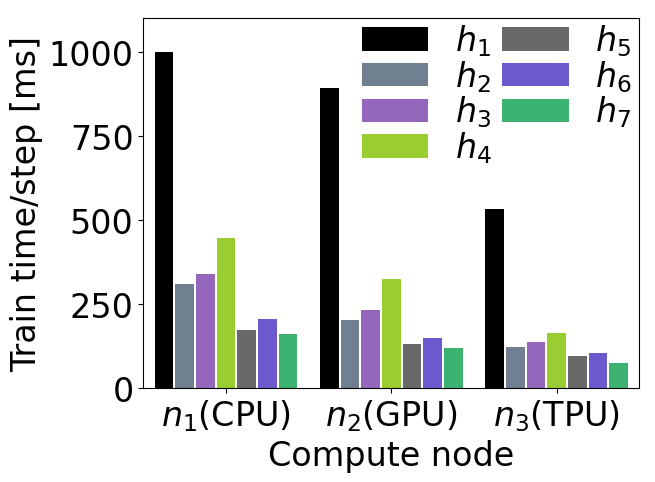}\vspace{-3mm}
    }\hspace{-3mm}  \subfigure[\vspace{-3mm}\label{f:train_time_app}]{      
    \includegraphics[width=0.48\columnwidth]{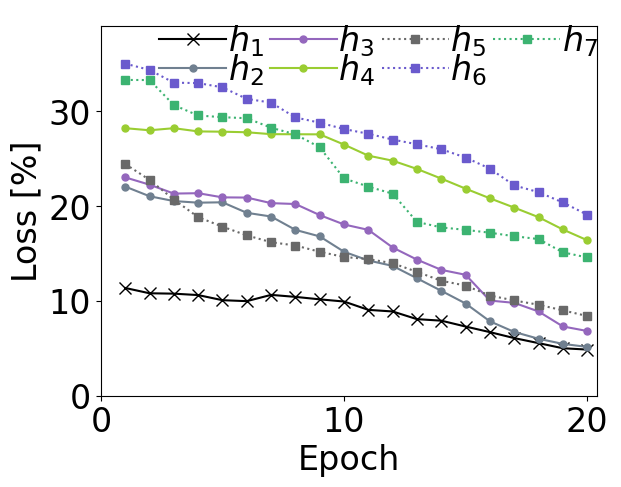}\vspace{-3mm} 
         }
         \vspace{-3mm}
         \caption{(a) Training time/step on compute nodes; and (b) Loss (MAPE) (config.: 20 epochs, 2000 steps/epoch) for applications listed in Table \ref{t:app}. \vspace{-3mm}}
 \label{f:res_app}
 \end{figure}

\begin{figure*}[htb]
    \centering     
         \includegraphics[width=0.2\textwidth]{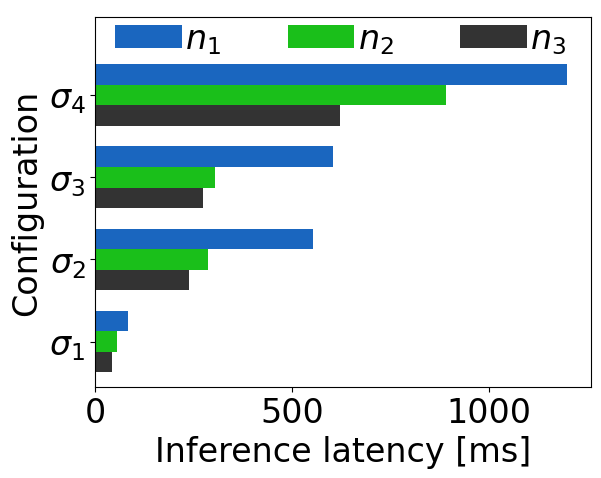} 
\hspace{2mm}
\includegraphics[width=0.2\textwidth]{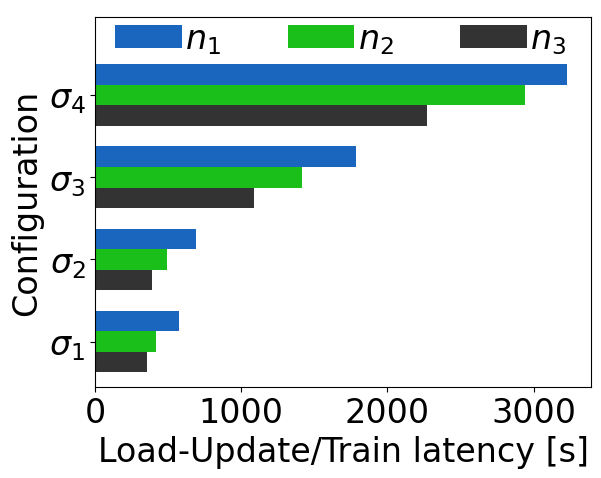}  
\hspace{2mm}
\includegraphics[width=0.2\textwidth]{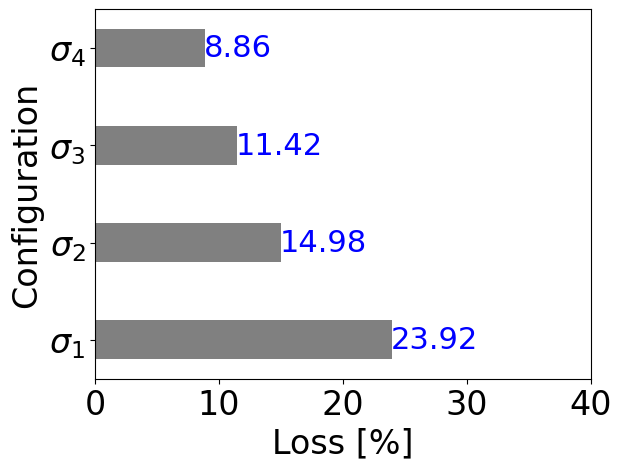}  
\hspace{2mm}
\includegraphics[width=0.2\textwidth]{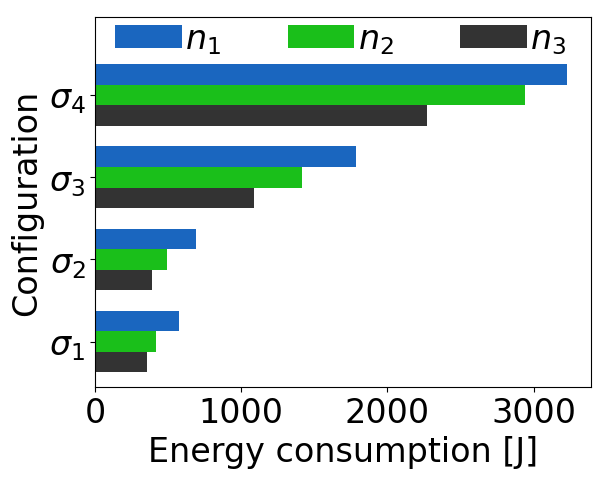} 
\vspace*{-2mm}
    \caption{Impact of the configurations ($\sigma_{1-4}$) execution by compute nodes ($n_{1}$:CPU, $n_2$:GPU, and $n_3$:TPU) listed in Table\,\ref{t:compute}: inference latency (left), load-update or train latency (center-left), loss (center-right) and energy consumption (right) for the $h_1$ application.\vspace*{-3mm}}
    
\label{f:res_config}
\end{figure*}
\begin{figure*}[htb]
    \centering     
         \includegraphics[width=0.41\textwidth]{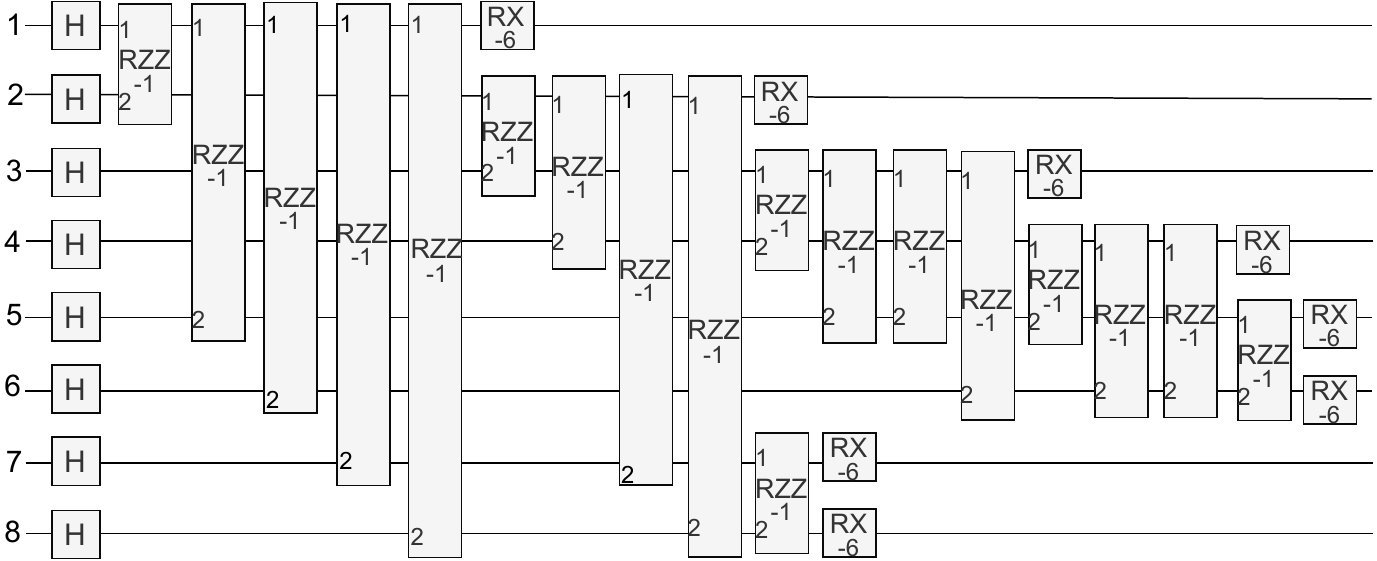} 
\hspace{2mm}
\includegraphics[width=0.21\textwidth]{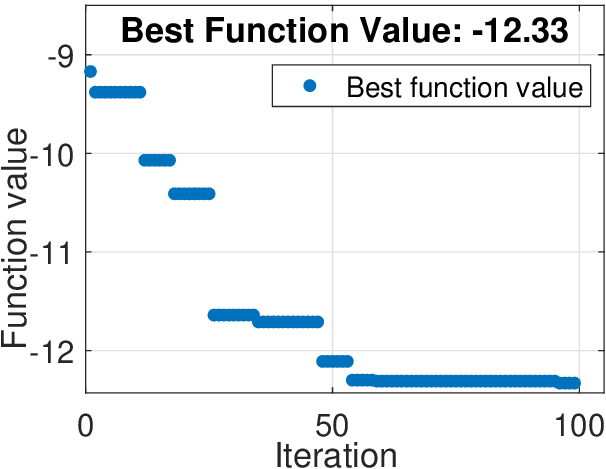}  
\hspace{2mm}
\includegraphics[width=0.21\textwidth]{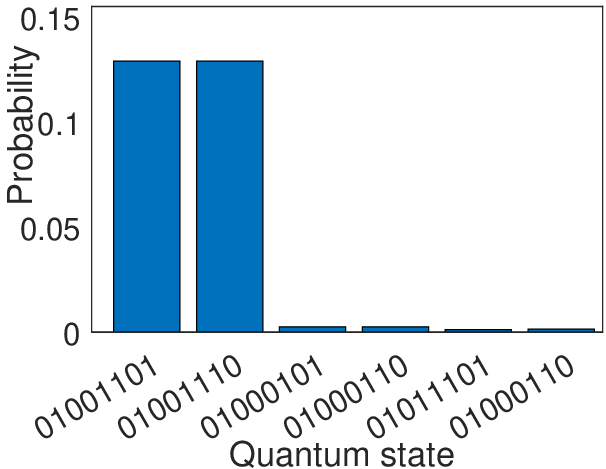}  
\vspace*{-3mm}
    \caption{Quantum circuit (left), iterative function value (center), and Quantum state probability (right) for the small-scale scenario shown in Fig. \ref{f:example}.\vspace*{-5mm}}
\label{f:quant_2}
\end{figure*}

\noindent
\textbf{Compute Nodes. } 
To evaluate the performance of the proposed QAG solution, we consider three types of compute nodes, CPU, GPU, and TPU~\cite{cpu_gpu_1}, respectively, associated with the values of computational capability in trillions of operations per second (TOPS) and power consumption in watt (W), listed in Table\,\ref{t:compute}. We monitor the processor usage, architecture information, and power usage with the help of cross-platform libraries and commands (lscpu, psutil) along with the TensorFlow profiler and NVIDIA Management Library (nvml). We recall that such parameters define the computing resource capacity of the vertices $n_k\in\Vc_3$. 

\noindent
{\bf Benchmarks.}  We compare QAG against:\\
\textbullet\,\emph{RouteNet-Fermi (RNF)}, \cite{routenetFermi}. 
We select~\cite{routenetFermi} because it is a recent state-of-the-art for GNN-based network modeling for all the considered applications to predict delay, jitter, and packet-loss. However, it has a fixed configuration for model training and a fixed set of pre-trained models for inference.\\
\textbullet\,\emph{Optimum (Opt),} obtained with brute-force exhaustive search.  


\subsection{Results and Discussions}

\noindent{\bf Latency, quality, and energy-consumption performance.}

We evaluate the inference time per sample and loss performance (on average for $>$4000 samples) when using a pre-trained model without any updates/training in Fig.\,\ref{f:res_infer}. We observe that the inference time per sample on the TPU is lesser than GPU, and it is lesser on GPU than on CPU. However, the quality of inference is very poor, i.e. inference loss is high ($>100 \%$ MAPE) for three applications ($h_4,h_6,$ and $h_7$) while it is still considerable ($>23 \%$ MAPE) for the other applications. We note that even though inference using a pre-trained model is faster having at most 143 ms on TPU for $h_4$ application but it is not a viable option due to an extremely poor inference quality (extremely high loss). Hence, loading a pre-trained model for inference might not always be suitable to meet the network modeling application requirements. 

Next, we evaluate the trade-off between training time (per step) and the quality of training configs. for the considered applications (Table\,\ref{t:app}) and compute nodes (Table\,\ref{t:compute}) in Fig.\,\ref{f:res_app}. The loss performance, measured by MAPE, improves (i.e., loss decreases) with an increasing number of epochs for all the applications, but the extent of decrease in the loss is different for each application. Even the per-step training time is lesser on TPU as compared to GPU and CPU. Hence, an application and compute platform aware adaptive configuration selection and deployment is necessary for efficient network modeling. 

We further evaluate the performance (energy consumption, latency, and quality) of four sample configurations, $\sigma_1=\{\text{ABILENE}, 1,1,0 \}$, $\sigma_2=\{\text{GEANT}, 1,5,1 \}$, $\sigma_3=\{\text{GEANT}, 10,50,1 \}$, and $\sigma_4=\{\text{GEANT}, 20,200,1 \}$, that are deployed on compute nodes listed in Table \ref{t:compute}, for the application $h_1$ in Fig.\ref{f:res_config}. This shows that a dynamic decision of inference or update-train after loading a pre-trained model on a chosen compute platform for a specific application has a direct impact on the performance (latency, quality, and energy). 

Fig.\ref{f:quant_2} shows the optimal quantum circuit, iterative objective function value, and the probability of the quantum state, for the small-scale example that is depicted in Fig.~\ref{f:example}. The objective function, \eqref{qaoa_eq}, converges to the minimum in less than $60$ iterations. Specifically, we observe that the highest probability quantum states, $\ket{01001101}$ and $\ket{01001110}$, correspond to the sub-graphs ($h_1, \sigma_1, \sigma_2, n_{1 \text{ or } 2}$) and ($h_2, \sigma_3, \sigma_4, n_{2 \text{ or } 1}$).

\begin{figure}[!htb]
\centering
\subfigure[$\tau^h_{\max}=$5\,s \label{f:energy_loss_small}]{
\includegraphics[width=0.4\columnwidth]{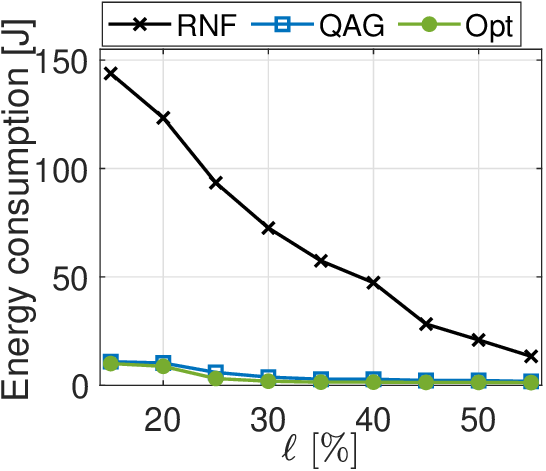}  
    }\hspace{-2mm}
\subfigure[$\ell^h_{\max}=$20$ \%$\label{f:energy_lat_small}]{   \includegraphics[width=0.4\columnwidth]{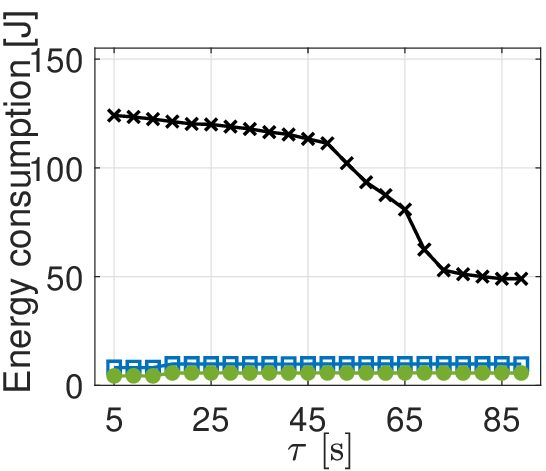}
         }
         \vspace{-4mm}
         \caption{Small scenario: Energy consumption as  target loss and latency vary.}
          \vspace{-4mm}
 \end{figure}
\begin{figure}[!htb]
\centering
\subfigure[$\tau^h_{\max}=$5\,s \label{f:churn_loss_small}]{
\includegraphics[width=0.4\columnwidth]{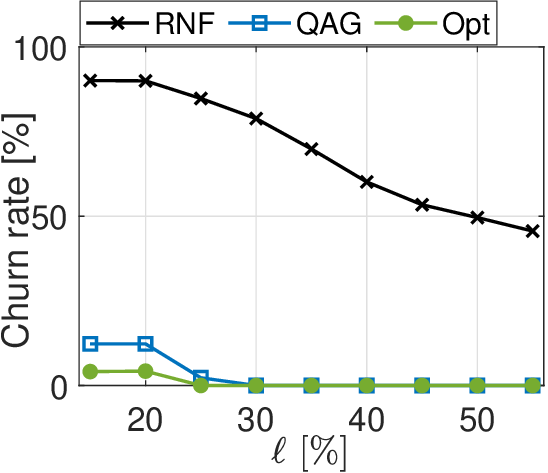}  
    }\hspace{-2mm}
\subfigure[$\ell^h_{\max}=$20$ \%$\label{f:churn_lat_small}]{   \includegraphics[width=0.4\columnwidth]{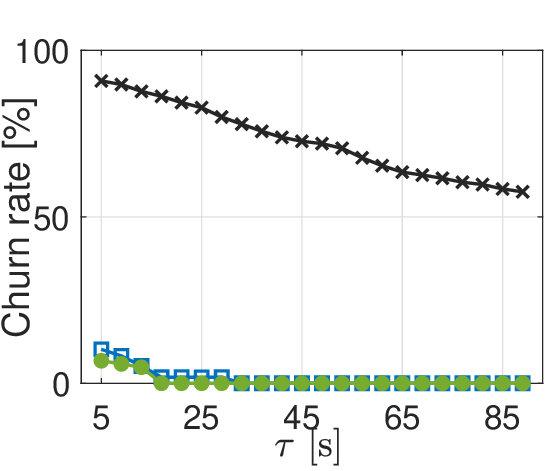}
         }
         \vspace{-4mm}
         \caption{Small scenario: Churn rate obtained as the target loss and latency vary.\vspace{-5mm}}
 \end{figure}
 \begin{figure}[!htb]
 \centering
\subfigure[$\tau^h_{\max}=$5\,s \label{f:energy_loss_large}]{
\includegraphics[width=0.4\columnwidth]{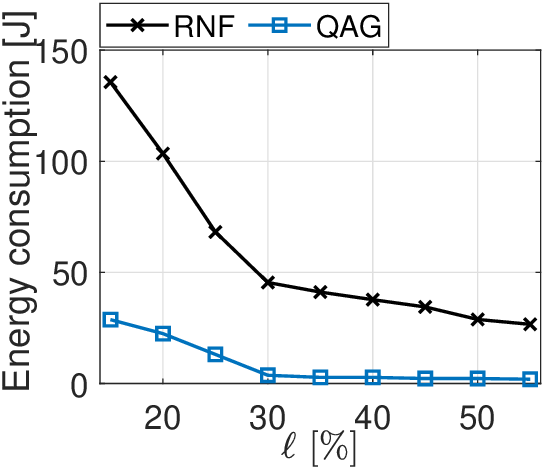}  
    }\hspace{-2mm}
\subfigure[$\ell^h_{\max}=$20$ \%$\label{f:energy_lat_large}]{   \includegraphics[width=0.4\columnwidth]{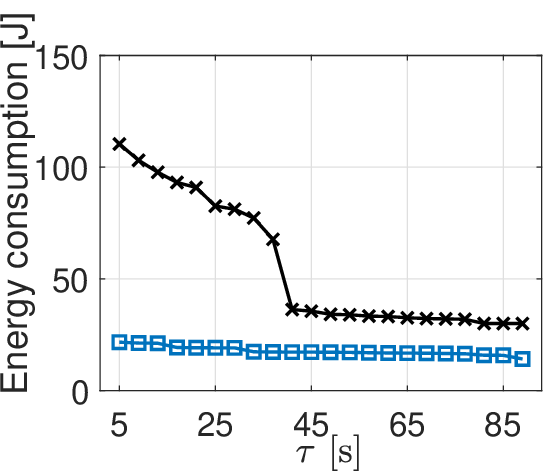}}
         \vspace{-4mm}
         \caption{Large scenario:\,Energy consumption as  target loss and latency vary.\vspace{-2mm}}
          \vspace{-2mm}
 \end{figure}
\begin{figure}[!htb]
\centering
\subfigure[$\tau^h_{\max}=$5\,s \label{f:churn_loss_large}]{
\includegraphics[width=0.4\columnwidth]{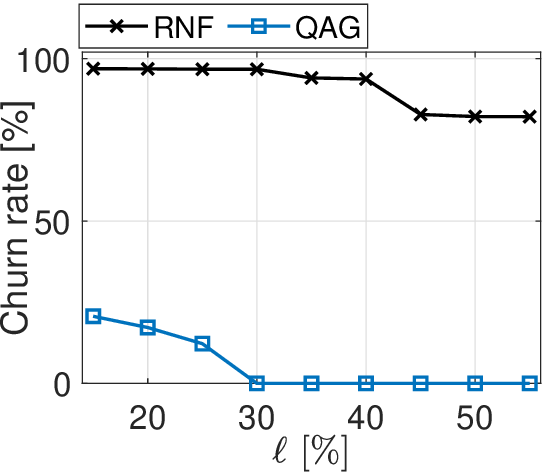}  
    }\hspace{-2mm}
\subfigure[$\ell^h_{\max}=$20$ \%$\label{f:churn_lat_large}]{   \includegraphics[width=0.4\columnwidth]{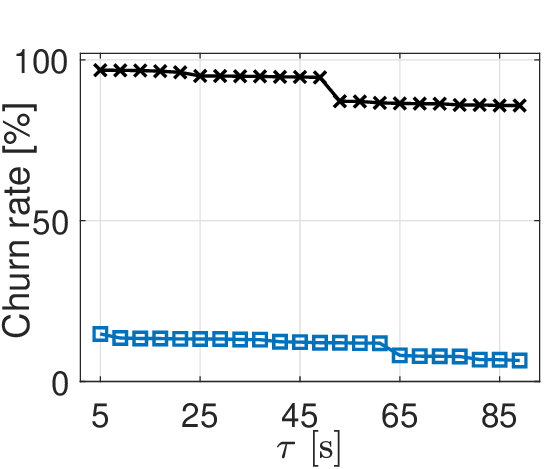}}
         \vspace{-4mm}
         \caption{Large scenario: Churn rate as the target loss and latency vary.\vspace{-5mm}}
 \end{figure}

\noindent{\bf Small-scale two-application scenario.}
We consider the small-scale two-application scenario similar to the example shown in Fig.~\ref{f:example} in order to evaluate the efficacy of our proposed QAG solution with respect to the Optimal solution ({\em{Opt}}). It  consists of two-applications, four configurations, and two compute nodes, that are uniformly randomly selected from the possible options for each. We evaluate the energy consumption and churn-rate (proportion of instances failing to meet the application requirements) performance over 10000 iterations with 95\% confidence interval, with varying latency and loss requirements for the applications.

Fig.\ref{f:energy_loss_small} and Fig.\ref{f:energy_lat_small} show the energy-efficiency, Fig.\ref{f:churn_loss_small} and Fig.\ref{f:churn_lat_small} show the churn-rate performance, of the RNF, QAG, and Opt schemes in the small-scale scenario. We observe that both the energy consumption and the churn rate reduce as the latency target or the loss target increases due to the increased number of feasible configurations meeting the requirements. The proposed QAG performance closes matches the Opt, and it outperforms the RNF scheme in all considered requirement (loss and latency) settings. Overall, QAG results in at least a 70\% lower energy consumption and a 60\% lower churn rate than the benchmark RNF scheme.

\noindent{\bf Large-scale seven-application scenario.}
We consider a large scale scenario consisting of the seven applications listed in Table~\ref{t:app}, twenty possible configurations for each application, and three compute nodes of each type that are listed in Table~\ref{t:compute}.
The optimal solution is infeasible to obtain for a large scenario consisting of many applications, possible configurations, and compute nodes. This is due to the extremely high complexity of the exhaustive search method in finding the optimal solution for a large scenario that results in a large tripartite graph consisting of many vertices ($N>10$). Hence, we perform the comparative energy consumption and churn-rate evaluation of the proposed QAG with respect to the state-of-the-art RNF scheme for varying latency and loss requirements of the considered applications.

For the large-scale scenario  with loss target, $\ell^h_{max}=20\%$, Fig.\ref{f:energy_loss_large} and Fig.\ref{f:churn_loss_large} show the energy-efficiency and churn-rate performance of the RNF and QAG schemes as the latency target varies. When we set the latency target, $\tau^h_{max}=$ 5s for the applications, Fig.\ref{f:energy_lat_large} and Fig.\ref{f:churn_lat_large} show the performance as the loss target varies. We observe that both the energy consumption and the churn rate reduces for QAG as the latency target or the loss target for the applications increase. This is due to the increased number of feasible configurations and need for lesser compute resources to meet the less stringent application requirements. The proposed QAG yields in atleast a 50\% lower energy consumption and a 80\% lower churn rate than the benchmark RNF scheme.


\section{Conclusions\label{sec:conclusions}}
We solved the problem of orchestrating the adaptive GNN-based network modeling in the mobile-edge-cloud continuum system in an energy-efficient way. The network modeling applications using the GNN-based models can have diverse latency and inference quality (loss) requirements, we meet these by dynamically selecting the efficient GNN-based model configuration and deploying it on the suitable network compute node. Our solution approach, QAG, leverages the tripartite graph representation of the system and applies a low-complexity quantum approximation optimization to find the energy-efficient orchestration solution for heterogeneous network modeling applications. Extensive evaluation with different, co-existing network modeling applications for a small-scale and large-scale scenario demonstrates that our solution performs very closely to the optimum and, compared to the existing alternative, it can reduce the energy consumption by more than 50\% while meeting the application requirements with at least a 60\% lower churn-rate.  

\vspace{-2mm}
\section*{Acknowledgement}
\small{This work was supported by the 5G Events Labs project funded by BPI France and the PEPR 5G project funded by the French National Research Agency (ANR).\vspace{-2mm}}

\bibliographystyle{IEEEtran}
\bibliography{main.bib}

\end{document}